# Phonon driven spin distribution due to the spin-Seebeck effect

C. M. Jaworski[1], J. Yang[2], S. Mack[3], D. D. Awschalom[3], R. C. Myers[2,4*], and J. P. Heremans[1,4*]

1. Department of Mechanical Engineering, The Ohio State University, Columbus, OH
2. Department of Materials Science and Engineering, The Ohio State University, Columbus, OH
3. Center for Spintronics and Quantum Computation, University of California, Santa Barbara, CA
4. Department of Physics, The Ohio State University, Columbus, OH

*To whom correspondence should be addressed. Email: heremans.1@osu.edu, myers.1079@osu.edu

**Here we report on measurements of the spin-Seebeck effect of GaMnAs over an extended temperature range alongside the thermal conductivity, specific heat, magnetization, and thermoelectric power. The amplitude of the spin-Seebeck effect in GaMnAs scales with the thermal conductivity of the GaAs substrate and the phonon-drag contribution to the thermoelectric power of the GaMnAs, demonstrating that phonons drive the spin redistribution. A phenomenological model involving phonon-magnon drag explains the spatial and temperature dependence of the measured spin distribution.**

The spin-Seebeck effect, consisting of a thermally generated spin redistribution, has been observed in spin-polarized metals[1], semiconductors[2] and insulators[3, 4, 5]. This effect joins other newly discovered spintronic phenomena, such as the spin Hall effect[6]. While the experimental evidence has been reproduced by several groups, recent theoretical developments[7, 8, 9] have not been able to explain simultaneously the combination of the persistence of the effect after the severance of electrical communication[2], the positional dependence, and the temperature dependence[1, 2, 3]. Here we provide a detailed characterization of the temperature dependence of the spin-Seebeck coefficient ($S_{xy}$), magnetization (M), and thermoelectric power (thermopower or $α_{xx}$) of the ferromagnet, alongside the thermal conductivity (κ) and specific heat ($C_p$) of the substrate. These measurements reveal a direct correlation with the amplitude and temperature dependence of the spin-Seebeck coefficient in multiple samples from which we conclude that the spin-Seebeck effect is driven by phonons. A simple phenomenological model involving magnon-phonon drag explains the general features of the temperature dependence and the spatial dependence of the spin-Seebeck effect in GaMnAs.

The spin-Seebeck effect produces a voltage across a platinum strip deposited on top of a ferromagnet in response to an applied temperature gradient. The sample geometry is shown in Fig 1 and is the same as Refs. [1, 2, 3, 10]. The Pt strips act as a spin-polarization detector based on the inverse spin Hall effect (ISHE)[11], and the measured voltage, or integrated electric field $E_{ISHE}$, is a measure of the vertical flux of spin-polarized particles that reach from the ferromagnetic layer into the Pt. $E_{ISHE}$ is linear with respect to the temperature gradient. Unlike classical transport coefficients, it is not uniform across a sample, but depends spatially on the position x around the center of the sample, following a $E_{ISHE} \propto \sinh(x/\lambda-b)$ relation with a characteristic length scale λ of millimeters. The spin-Seebeck effect persists without electrical communication between ends of the sample.

Here, the samples consist of a thin film of ferromagnetic $Ga_{1-s}Mn_sAs_1$ grown on semi-insulating GaAs with s=0.158 (Fig. 2) and s=0.16 (Fig. 3). One end of the sample is heated (x= -L/2) the other cooled (x= +L/2) thus creating a longitudinal gradient $\nabla T_x$. The voltage developed by the Pt strips along the *y* direction by spin-polarization in GaMnAs is measured. A spin-current traveling (along *z*) from the ferromagnet into the Pt is scattered differentially depending on the spin polarization and generates the electric field $E_{ISHE}$ that is mutually orthogonal to the spin current (along z) and spin polarization (along x) directions[11]. We measure $E_{ISHE}$ as a voltage $V_y$ across the Pt strips and sweep magnetic field (along x) in a hysteric manner to realize reversals in direction of $E_{ISHE}$ at the coercive field as a measurable $\Delta V_y$. We subsequently subtract out residual voltages and EMF pickup from sweeping magnetic field, thus centering the $V_y$ signal around zero volts. The linearity of $\Delta V_y$ vs. $\Delta T_x$ has been shown elsewhere[2], and enables us to define the spin-Seebeck coefficient $S_{xy}$ as half of $\Delta V_y$ normalized by width of sample and distance between the thermometry, $S_{xy} \equiv \frac{E_y}{\nabla_x T} = \frac{L \Delta V_y}{2w \Delta T_x}$. The positional dependence of $S_{xy}(x) \propto \sinh(x/\lambda - b)$ has also been shown in Ref. [2]; here we concentrate on the magnitude $S_{xy}$ on a contact near x=-L/2, where it is maximal, thus allowing for the best signal to noise ratio. Given the definitions above, $S_{xy}(-L/2)<0$, and $S_{xy}(+L/2)>0$, and preliminary data on other ferromagnets[2, 12], show that these signs are material-dependent. The experimental methods and sample preparation are the same as in Ref. [2] but in a modified cryostat (Thermal Transport Option in a Quantum Design Physical Properties Measurement System) that enables a better heat sinking of the heat flux applied to the sample during the measurements, thus allowing us to extend our measurement temperature range of $S_{xy}$ downward. We measured thermal conductivity and thermopower using a classical heater-and-sink method. Specific heat is measured using a quasi-adiabatic platform calorimeter; magnetic moment using a SQUID magnetometer. All reported data were measured on two samples $4 \times 12$ mm$^2$ of 30 nm thick $Ga_{.0842}Mn_{0.158}As$ (Fig. 2) or $5 \times 15.5$ mm$^2$ 100 nm thick $Ga_{.084}Mn_{0.16}As$ (Fig. 3) with a magnetic easy axis along [$\bar{1}10$] grown on 0.5 mm thick semi-insulating GaAs by molecular beam epitaxy[2, 13]. The error in $S_{xy}$ is estimated by dividing the RMS of the noise on the voltmeter (Keithley 2182A) by $\Delta T_x$ whereas other error estimates are provided by the instrumentation software.

In Fig. 2 we show $S_{xy}(T)$, the magnetization, and the thermopower of one GaMnAs sample (green stars) alongside the sample substrate's thermal conductivity. The substrate's specific heat is given in the inset to Fig. 2a, and portions of $S_{xy}$ and $\kappa$ data are fitted to power laws. Thermopower of another piece of the similar sample (purple circles) extended to lower temperature is included in Fig. 2c to illustrate the phonon-drag effect in thermopower. The measured $\kappa$ and $C_p$ are actually the sum of the contributions of both the GaAs substrate and the GaMnAs film, but in practice the 0.5 mm-thick substrate phonons dominate both because the film is only 30 nm thick. The magnetization shows a ferromagnetic behavior with $T_c \sim 135$ K. We note four distinct temperature regimes, separated by vertical dashed lines, in the spin-Seebeck effect in Fig. 2.

(1) Above the Curie temperature ($T_c$) of 135 K the sample is a paramagnet, and $S_{xy} = 0$.

(2) Between 85-135 K both M and $S_{xy}$ show an order parameter behavior $(T_c-T)^{-\gamma}$. $S_{xy}$ increases in magnitude with decreasing T to $| S_{xy} | \sim 0.25$ μV/K, demonstrating the dependence on the GaMnAs magnetization. The thermal conductivity of the substrate increases with decreasing T but much more slowly than either M or $S_{xy}$. The thermopower $\alpha_{xx}$ is proportional to T as expected in a degenerately-doped semiconductor.

(3) Between 85 K and the maximum in both $S_{xy}$ and $\kappa$ at 35 K, the magnetization varies more slowly with temperature. The thermal conductivity of the substrate increases with decreasing T as a $T^{-1}$ law, characteristic of anharmonic phonon-phonon Umklapp scattering, then reaches a maximum at 35 K. The spin-Seebeck signal $| S_{xy} |$ follows $\kappa$ closely up to a peak of ~1.05 µV/K at 35 K and both $S_{xy}$ and κ peak at the same T. The thermopower $α_{xx}$ departs strongly from the $T^1$ and the difference forms a peak also slightly below 35 K indicative of phonon-electron drag.

(4) Below 35 K, there is a sharp decrease in $S_{xy}$ with decreasing T following a $T^{3/2}$ law, with | $S_{xy}$ | reaching ~0.4 µV/K at 15 K on the sample in Fig. 2. This $T^{3/2}$ behavior is the temperature-dependence of a magnon specific heat[14]. The specific heat of the substrate follows a $T^3$ law consistent with the Debye model, though a slightly slower slope appears at the lowest temperatures, which may be due to an incipient electron ($C_p \propto T$) or a magnon ($C_p \propto T^{3/2}$) contribution. The thermal conductivity also follows a $T^3$ law down to 10 K: this is understood by realizing that the phonon mean free path Λ now is a constant of the order of the sample thickness, and κ = 1/3 $C_p$ v Λ, where v is the sound velocity. Again, there is excess conduction at the lowest temperature, possibly due to the same cause as in $C_p$. The phonon-drag effect on the thermopower $α_{xx}$ decreases at lower T where it diminishes to zero at 0 K.

The above trends are reproduced on a second sample grown on a higher quality GaAs substrate, and consequently ten times higher peak thermal conductivity (Fig. 3). The maximum in κ(T) is now lowered to ~10-15 K; high thermal conductance makes accurate measurement near the maximum difficult. The spin-Seebeck coefficient now peaks at 10 K instead of 35 K in the previous sample (Fig. 2). As the Pt strips on the two different samples are not exactly at the same *x* position nor do they have exactly the same thickness nor coupling to the GaMnAs, we refrain from a direct comparison. We do note that the sample in Fig. 2 at 50 K | $S_{xy}$ | ~0.75 µV/K which peaks at ~1.05 µV/K at 35 K while the sample in Fig. 3 at 50 K | $S_{xy}$ | ~0.2 µV/K which peaks at ~ 4 µV/K at 10 K. The magnitude scales roughly with the absolute value of the substrate

thermal conductivity. The smaller value of $S_{xy}$ at 50 K for Fig. 3 is because the Pt strip was located further away from the hot end of the sample (*x= -L/2*).   The phonon-drag peak in $\alpha_{xx}$ is now a prominent feature with a maximum again at ~10-15 K.  This reveals that the amplitude of the spin-Seebeck effect in GaMnAs scales with the thermal conductivity of the substrate as well as with the intensity of phonon-electron drag in the thermopower as a function both of substrate condition and of absolute temperature.

The observation that spin-Seebeck scales with thermal conductivity and phonon-electron drag is consistent with recent publications invoking magnon-phonon drag as a mechanism at least contributing to the spin-Seebeck signal of the ferromagnetic insulator YIG[7, 8, 9], although equation (3) in Ref. [9] does not fit our $S_{xy}$ data.. First, we review the mechanism behind the phonon-electron drag contribution to the thermopower $\alpha_{xx}$. At higher temperatures, the classical diffusive thermopower is governed by the Boltzmann equation. In this regime, and for degenerately-doped semiconductors, $\alpha_{xx} \propto T$ as observed above 90 K in Fig. 2c.  Here electrons and phonons are constantly being brought back to mutual equilibrium by collisions, and the temperature gradient creates only a small perturbation of the electron equilibrium distribution function. When electron-phonon interactions dominate over other scattering mechanisms of both electrons and phonons, the phonon drag thermopower adds to the diffusive thermopower in the form of a peak at a certain temperature. When enough phonons interact with electrons, rather than with other phonons or impurities and defects, they impart momentum to the electrons along $\nabla T$ and move the electron distribution function away from equilibrium, resulting in an extra phonon-drag thermopower that can be orders of magnitude larger than the diffusive thermopower. The amplitude of this effect scales with the ratio between phonon/electron and phonon/phonon or phonon/defects interaction cross-sections, as well as with the density of phonons available to interact with the electrons. Therefore, in metals the phonon-drag thermopower peaks at a temperature close to the maximum in lattice thermal conductivity. Above that maximum, Umklapp processes compete with phonon-electron interactions to bring the phonons back to equilibrium, while below that temperature the number of phonons decreases following the Debye specific heat. The situation in degenerately-doped semiconductors is only slightly different: as observed in Figs. 2 and 3 the temperature of the maximum in the phonon drag contribution can differ somewhat from that of the maximum in the lattice thermal conductivity because it depends on the cross-section of the Fermi surface and the number of phonons that can interact with electrons on that surface. Comparing Figs. 2 and 3 illustrates that the lower the phonon-phonon and phonon-defect interactions, the higher the relative magnitude of phonon-electron drag and the concomitant thermopower $\alpha_{xx}$.

In Fig. 1, we offer a qualitative outline of the role of substrate phonons in the thermally-induced spin distribution. At least in the case of ferromagnetic insulators[7], the driving force for phonon-magnon drag is the difference between the temperature of the magnons $T_M(x)$ in the ferromagnetic film and that of the phonons $T_P(x)$ in the GaAs substrate and in the film (Fig. 1a); given the similarities between our observation of the spin-Seebeck effect in GaMnAs and that in YIG[3], we assume a similar case to hold here. The drag force only arises in the presence of a temperature gradient, which imparts an excess momentum to the phonons; in the absence of a gradient, $T_M(x) = T_P(x)$ at all x. In the presence of a gradient, $T_P(x)$ follows a linear profile between the hot and the cold end of the sample (Fig. 1a). We assume that the temperature baths at the ends of the sample, i.e. the heater at the hot end and the heat sink at the cold end, only connect to the phonons, and that the magnons that interact with phonons can tunnel large distances and thus cross the 300 μm gaps in the film (Fig. 3 of Ref. [2]) resulting in an uninterrupted $T_M$ profile (Fig. 1a). The difference $\Delta T_M(x) = T_P(x) - T_M(x)$ between the two is calculated to follow a $\sinh(x/\lambda)$[14] law (Fig. 1b), and mirrors the observed spatial dependence of $S_{xy}(x)$ [2]. Near the center of the sample, $\Delta T_M(x\sim 0)$ must be zero; the exact location depends on the thermal symmetry of the setup, in particular of the coupling to the reservoirs at the ends of the sample, and can be slightly offset from x = 0. At the hot end of the sample, $\Delta T_M(x < 0) > 0$, phonon-magnon drag tends to heat up the magnons; the reverse holds at the cold end where the phonons cool the magnons, $\Delta T_M(x > 0) < 0$.

In the classical picture, magnons are perturbations of the ferromagnetic spins residing around Mn ions. The local moments fill precession cones (Fig. 1), and the average magnetic moment $M_x(T, H_x)$ is their projection along *x*. The zeroth order effect of the temperature gradient on these moments is via the temperature dependence of $M_x(T_M(x))$ as seen in Fig. 2a. Here, we envision a more dominant phonon-magnon drag mechanism. At the hot end of the sample, phonon drag heats up the magnons above thermal equilibrium, thus decreasing the average moment by a quantity $\Delta M_x (x < 0) < 0$ (Fig. 1c). At the cold end of the sample, the effect of drag is to cool magnons, increasing their average moment by $\Delta M_x(x > 0) > 0$. Thus $\Delta M_x(x) \sim \Delta T_M(x) \sim \sinh(x/\lambda)$. The dependence of $\Delta M_x$ on substrate $\kappa$ and on T is a function of the intensity of the phonon-magnon drag. As with phonon/electron drag, $\Delta M_x$ will therefore depend on the density of dragging phonons, and on the ratio of phonon/magnon to phonon/phonon and phonon/impurity interaction cross-sections. This is consistent with the observation that $S_{xy}(T) \sim \kappa(T)$ for T above the maximum in $\kappa$ for both samples (Figs. 2 and 3). At T < 35 K for the sample in Fig. 2, where $S_{xy}$ and $\kappa$ have a maximum, the magnon specific heat scales with $T^{3/2}$, rather than the $T^3$ for the phonons, and this is again consistent with the observed slope of $S_{xy}$(T < 35K) (Fig. 2b). Similar behavior is observed for the sample in Fig. 3, where the maximum now appears at a lower temperature of 15 K.

Next we consider the effect of $\Delta M_x(x)$ on the distribution of spin-polarized electrons. In GaMnAs, the charge carriers are highly spin-polarized (85%) holes[15]. The fact that a true zero (not an offset) is measured[2] for $S_{xy}$(x~0) near the middle of the sample indicates that spin polarized holes are not simply thermally diffusing into Pt since otherwise we would measure a signal even near x = 0 due to the inherent spin polarization of the Fermi carriers. The fact that a spin current in Pt is generated only in regions where there is a non-equilibrium spin distribution suggests that the two are related. Perhaps the non-equilibirum spin distribution (Fig. 1) is maintained by a continuous transfer of angular momentum to the magnons, thus requiring, by conservation, a flow of angular momentum in the form of a balancing spin-current. The spin-current then generates a voltage due to ISHE either in the Pt transducers or in the GaMnAs layer itself (Fig. 5 of Ref. [2]).

In conclusion, we have shown a clear correlation between the spin-Seebeck effect in GaMnAs, and the magnetization, substrate thermal conductivity, and phonon-drag thermopower. The effect of M(T) or κ(T) dominates in different temperature ranges, with M(T) having the greatest effect above 85 K, and κ(T) below. The scaling holds for the dependence of the effects on both temperature and sample crystalline quality. The thermodynamic coupling of spins and phonons, shown here, opens opportunities for fundamentally new spin-caloric concepts, either in reversible thermodynamics or in transport. This may lead to phonon engineering of spin-based devices, where heat transfer can be integrated with magnetic functionality and may result in fundamentally new applications, like spin-based cooling, or magnetically sensitive thermoelectrics and bolometers.

The authors thank E. Johnston-Halperin and K. J. Wickey for useful discussions. This work was supported in parts by the Center for Emergent Materials at the Ohio State University, an NSF MRSEC (Award Number DMR-0820414), NSF (including Award Number CBET-0754023), the ONR, the Ohio Eminent Scholar Discretionary Fund, and The Ohio State University Institute for Materials Research.

**Figure Captions**

FIG. 1 (color). A brief explanation of the spin-Seebeck effect. Top schematic shows the experimental setup (not to scale). (a) Temperature profile of phonons and magnons. (b) The temperature difference between phonons and magnons act as a driving force. Cartoon showing the effect of phonon drag on the magnetic moment $M_x$ in each region. (c) Change in $M_x$ due to phonon drag across the temperature gradient. (d) Schematic hysteresis loops representing the transverse voltage detected on the Pt strips in each region.

FIG. 2 (color). Thermal properties of one GaMnAs/GaAs sample versus temperature. (a) Magnetization of the GaMnAs and substrate (GaAs) thermal conductivity. Fits shown exhibit the following quality: $a+bT^3$ yields $R^2 = 0.985$, $T^{-1}$ yields $R^2 = 0.986$, $T_c-T^{-\gamma}$ yields $R^2 = 0.993$. The inset shows the substrate specific heat. (b) Spin-Seebeck coefficient $S_{xy}$ of the GaMnAs. Fits shown exhibit the following quality: $a+bT^{1.5}$ yields $R^2 = 0.954$, $a+bT^{-1}$ yields $R^2 = 0.917$, $T_c-T^{-\gamma}$ yields $R^2 = 0.953$. (c) Thermopower $\alpha_{xx}$ of the sample (green stars) and of a similar sample (purple dots). The vertical dashed lines divide the figure into four temperature ranges.

FIG. 3 (color). Thermal properties of a second GaMnAs sample grown on a substrate with higher thermal conductivity. (a) $S_{xy}$ and $\kappa$, and (b) $\alpha_{xx}$ and M as a function of temperature.

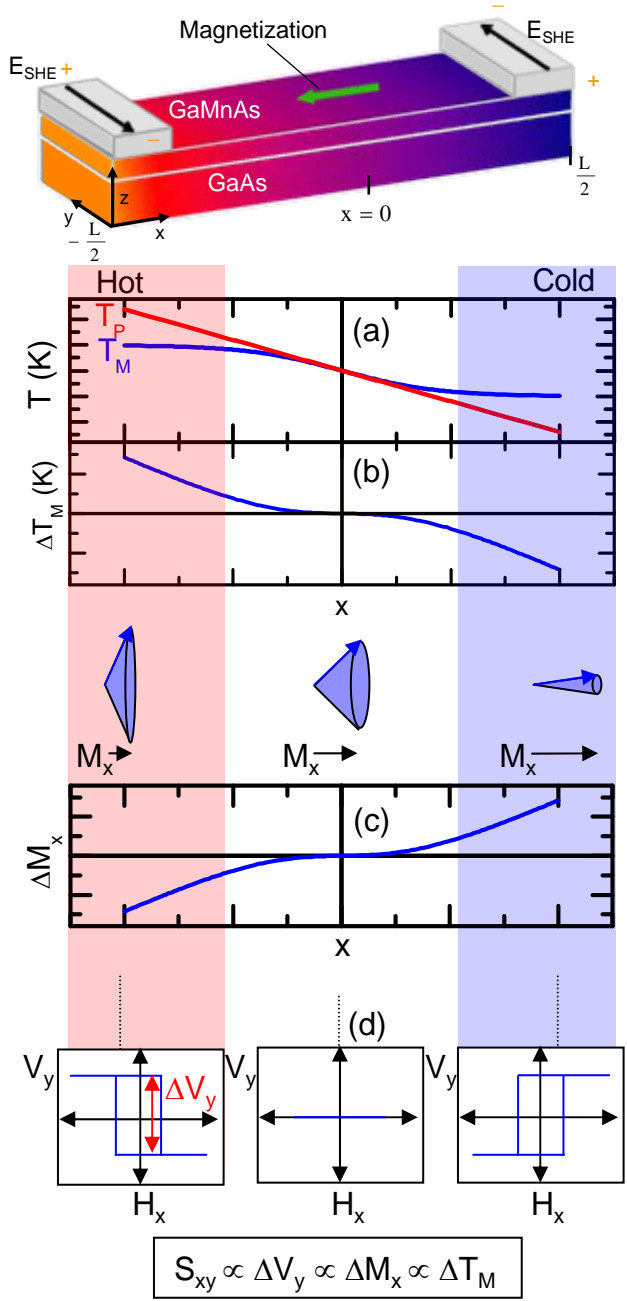

Fig. 1, Jaworski et al.

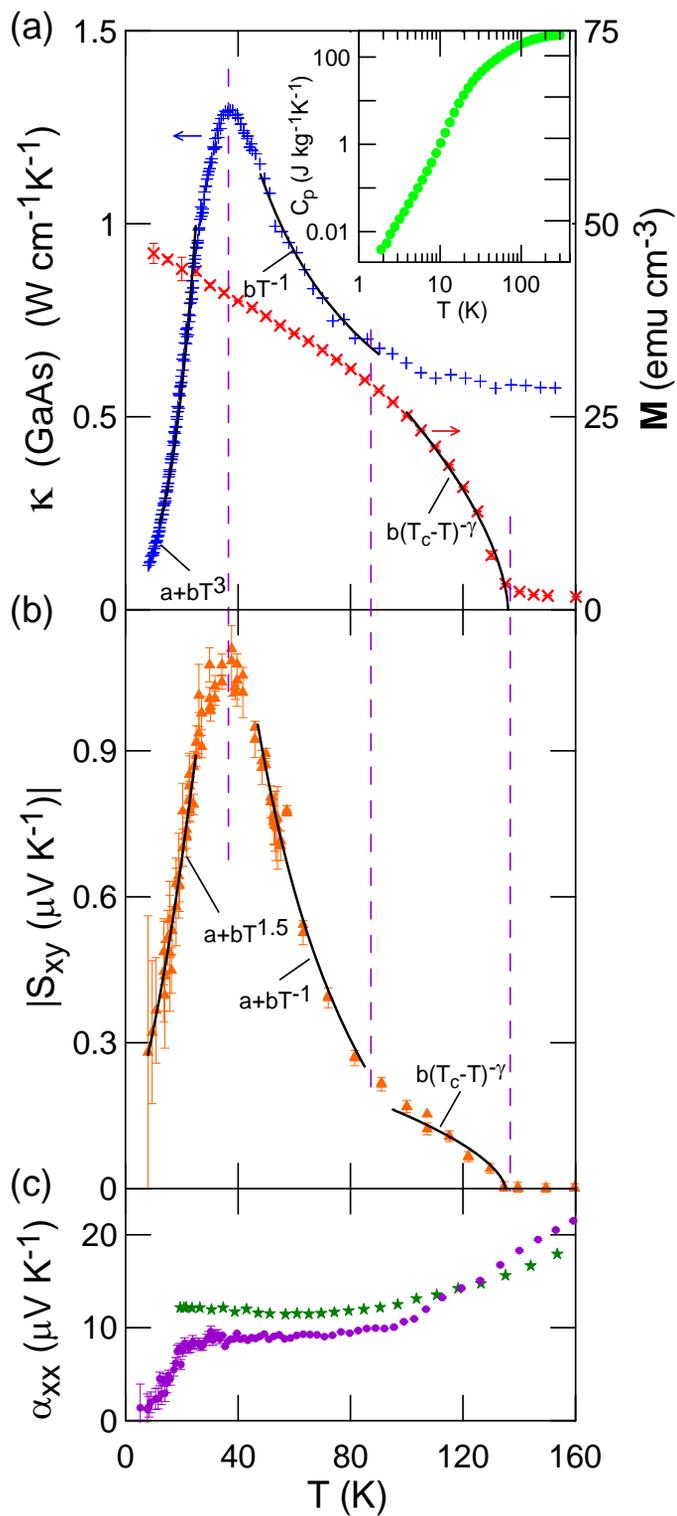

Fig. 2, Jaworski et al.

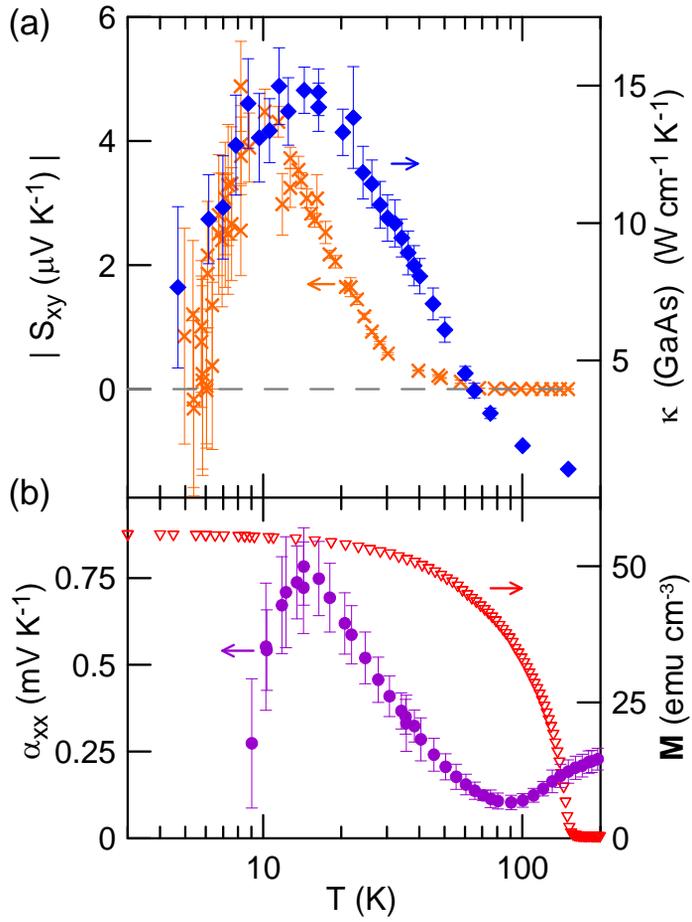

Fig. 3, Jaworski et al.